\documentclass{PoS}

\usepackage{mathtools}
\usepackage{array,contour}

\DeclareSymbolFont{myletters}{OML}{cmmi}{m}{n}
\DeclareMathSymbol{\varlambda}{\mathord}{myletters}{"15}
\contourlength{0.0em}

\newcommand{\doi}[1]{https://doi.org/#1}
\newcommand{\ie}{\textit{i.e.}}
\newcommand{\eg}{\textit{e.g.}}
\newcommand{\dd}{\mathop{}\!d}
\newcommand{\He}[1]{\ensuremath{{}^{#1}\mathrm{He}}}

\title{Time dependence of the p/He ratio in cosmic rays according to the force-field approximation}

\ShortTitle{Time dependence of the p/He ratio according to the force-field approximation}

\author{\speaker{Claudio Corti}\\
   University of Hawaii at Manoa\\
   E-mail: \email{corti@hawaii.edu}}
\author{Veronica Bindi\\
   University of Hawaii at Manoa}
\author{Cristina Consolandi\\
   University of Hawaii at Manoa}
\author{Christopher Freeman\\
   University of Hawaii at Manoa}
\author{Andrew Kuhlman\\
   University of Hawaii at Manoa}
\author{Cristopher Light\\
   University of Hawaii at Manoa}
\author{Matteo Palermo\\
   University of Hawaii at Manoa}
\author{Siqi Wang\\
   University of Hawaii at Manoa}

\abstract{We study the predictions for the p/He ratio in galactic cosmic rays according to the force-field approximation.
The dependence of the time variation of p/He on the local interstellar spectrum (LIS) shape and on the mass-to-charge ratio, $A/Z$, is analyzed in detail.
We find that, depending on the rigidity range and the sign of the spectral index of the p/He LIS ratio, the p/He time variation can be correlated or anti-correlated with the phase of the solar cycle.
We show that the $A/Z$ dependence is the most probable cause for the p/He decrease recently observed by AMS-02 after 2015 between 2 and 3 GV.}

\FullConference{36th International Cosmic Ray Conference -ICRC2019-\\
		July 24th - August 1st, 2019\\
		Madison, WI, U.S.A.}

\begin{document}

\section{Introduction}
The propagation of galactic cosmic rays (GCRs) in the heliosphere is affected by the heliospheric magnetic field (HMF) embedded in the solar wind \cite{bib:parker58:solar-wind,bib:parker65:modulation}.
GCRs are advected away by the solar wind, diffuse on the irregularities of the HMF, drift along the HMF gradients, curvature and the neutral heliospheric current sheet, and lose or gain energy adiabatically due to the solar wind expansion or contraction \cite{bib:potgieter13:solar-modulation}.
The Parker equation describes the transport of GCRs in the heliosphere \cite{bib:parker65:modulation}:
\begin{equation} \label{eqn:numerical-model:parker}
\frac{\partial f}{\partial t} + \mathbf{V}_{sw} \cdot \boldsymbol{\nabla}f - \boldsymbol{\nabla} \cdot \left( \boldsymbol{\mathsf{K}} \boldsymbol{\nabla}f \right) - \frac{\boldsymbol{\nabla} \cdot \mathbf{V}_{sw}}{3} \frac{\partial f}{\partial \mathrm{ln} R} = 0,
\end{equation}
where $f(\mathbf{r},R)$ is the omni-directional GCR distribution function at position $\mathbf{r}$ and rigidity $R$, $\mathbf{V}_{sw}$ is the solar wind speed, and $\boldsymbol{\mathsf{K}}$ is the diffusion tensor, which describes drifts and diffusion parallel and perpendicular to the average HMF direction.
The diffusion tensor is defined as $\boldsymbol{\mathsf{K}} =  \frac{1}{3}\beta$\contour*{black}{$\varlambda$}, where $\beta = v/c$ is the particle velocity divided by the speed of light, and \contour*{black}{$\varlambda$} is the mean free path tensor, related to the turbulent properties of the HMF \cite{bib:jokipii71:modulation}.
A general result of turbulence theory is that the drift and the diffusion mean free paths depend only on the particle rigidity $R$, so that \contour*{black}{$\varlambda$} is the same for all GCR nuclei.

\begin{figure}[h!]
   \centering
   \includegraphics[width=\textwidth]{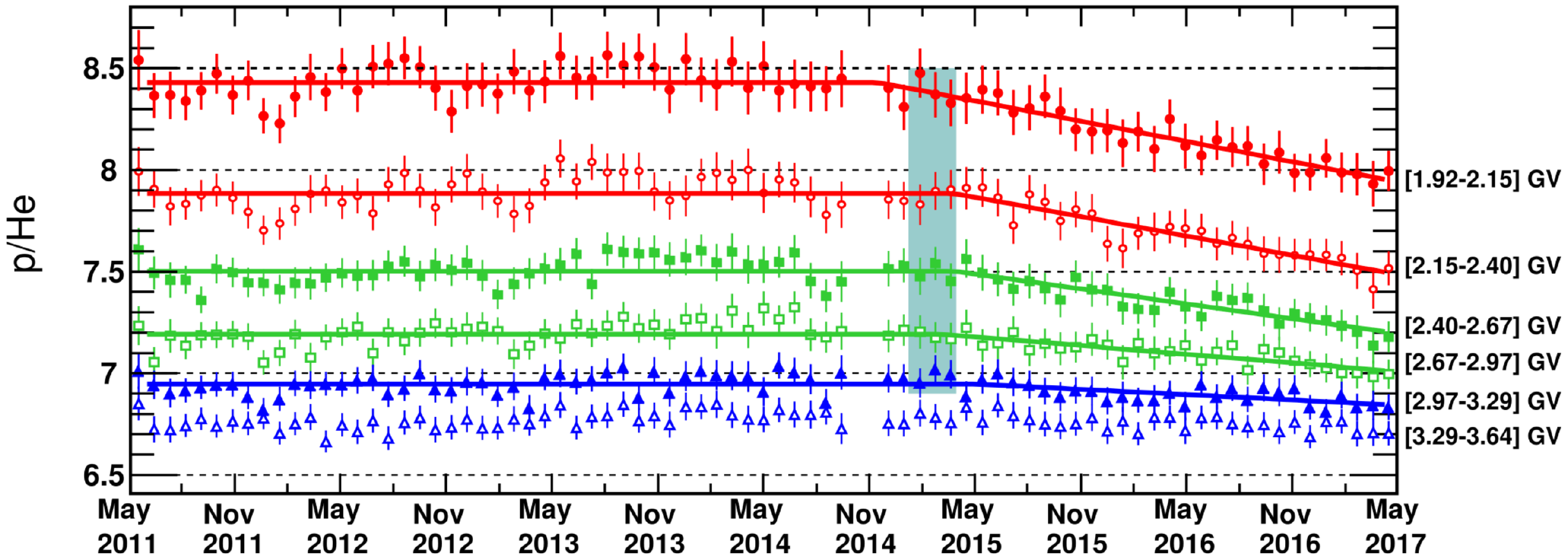}
   \caption{
      Time variation of the p/He flux ratio as measured by AMS for different rigidities (colored markers).
      The colored solid lines are the broken line fits performed by the AMS collaboration to test for the time dependence of the ratio, while the vertical box delimits the best-fit period of the beginning of the decrease.
   }
   \label{fig:ams-p-He}
\end{figure}

Recently, the Alpha Magnetic Spectrometer (AMS) experiment on board the International Space Station measured the time variation of GCR proton and helium fluxes, between May 2011 and May 2017, at monthly time resolution \cite{bib:aguilar18:ams-monthly-phe}.
The p/He flux ratio, as seen in Figure \ref{fig:ams-p-He},
has a clear long term trend in time below 3 GV: it remains flat until March 2015, and then it decreases by about 5\% around 2 GV in the next two years.

\emph{Tomassetti et al} \cite{bib:tomassetti18:model} and \emph{Corti et al} \cite{bib:corti19:model} reproduced the AMS observations using, respectively, a one-dimensional and a three-dimensional numerical model to solve the Parker equation, showing independently that the decrease in the p/He ratio over time is due to the difference in the mass-to-charge ratio, $A/Z$, between p and He.
The $A/Z$ dependence on solar modulation enters the Parker equation via the $\beta$ in the diffusion tensor.
On the other hand, \emph{Gieseler et al} \cite{bib:gieseler17:modffa} suggested that time variations in the flux ratio of two GCR species might be due to the different spectral slopes of their local interstellar spectrum (LIS). Different LIS lead to different solar modulation, as the LIS represents the boundary condition needed to solve the Parker equation.

In this work, we analyze the time variation of the flux ratio of two GCR species in the framework of the force-field approximation, which allows to explicitly see the dependence on the LIS and $A/Z$ in the analytical solution of the simplified one-dimensional Parker equation.
Although the force-field approximation is known not to be able to accurately reproduce solar modulated fluxes at Earth, especially during the solar maximum \cite{bib:corti19:ffa-validity}, the discrepancies mostly cancel out when taking the ratio of the fluxes, so that the conclusions can be safely applied to the study of the flux ratio of any two GCR species, and in particular to the p/He ratio observed by AMS.

\section{Flux ratio of two GCR species according to the force-field approximation}
\emph{Gleeson \& Axford} \cite{bib:gleeson68:ffa} derived a steady-state spherically symmetric analytical solution for the flux as function of kinetic energy, $J(T)$, under the assumptions that the convective and diffusive flows are equal, and that the diffusion coefficient $k(R) \propto \beta R$: $J(T) = R^{2}/R_{L}^{2} \ J_{L}(T_{L})$.
$J_{L}$ is the LIS (the flux at the heliopause), $T_{L} = T + Z\phi$ is the kinetic energy at the heliopause, $R_{L} = \sqrt{T_{L}(T_{L}+2Am)}$ the corresponding rigidity at the heliopause, $m$ is the proton mass, and $\phi$ is the so-called modulation potential, whose value depends on the phase of the solar cycle.

In order to describe the flux ratio of two species, $p_{1}$ and $p_{2}$, at a given rigidity, we must first rewrite the force-field solution in terms of $J(R) = Z\beta(R) J(T)$, the flux as function of rigidity:
\begin{equation} \label{eqn:ffa-rig}
J(R) = \frac{\beta(R)}{\beta(R_{L})} \ \frac{R^{2}}{R_{L}^{2}} \ J(R_{L}).
\end{equation}
Note that $\beta$, the Jacobian factor of the kinetic energy to rigidity conversion, does not simplify, as it is computed at different rigidities.
If we express $R_{L}$ as function of the measured rigidity $R$, $R_{L} = \sqrt{R^{2} + \phi^{2} + 2\phi R\sqrt{1 + (Am/ZR)^{2}}}$, and expand $\beta(R) = ZR/\sqrt{(ZR)^{2} + (Am)^{2}}$ and $\beta(R_{L}) = ZR_{L}/[\sqrt{(ZR)^{2} + (Am)^{2}} + Z\phi]$, then we find that the $p_{1}/p_{2}$ flux ratio is:
\begin{align}
   \frac{J_{1}(R)}{J_{2}(R)} & = 
   \left[ \frac{1 + \dfrac{\phi^{2}}{R^{2}} + 2\dfrac{\phi}{R}\sqrt{1 + \left(\dfrac{A_{2}m}{Z_{2}R}\right)^{2}}}{1 + \dfrac{\phi^{2}}{R^{2}} + 2\dfrac{\phi}{R}\sqrt{1 + \left(\dfrac{A_{1}m}{Z_{1}R}\right)^{2}}} \right]^{3/2} 
   \frac{\sqrt{1 + \left(\dfrac{A_{1}m}{Z_{1}R}\right)^{2}} + \dfrac{\phi}{R}}{\sqrt{1 + \left(\dfrac{A_{2}m}{Z_{2}R}\right)^{2}} + \dfrac{\phi}{R}} \
   \sqrt{\frac{1 + \left(\dfrac{A_{2}m}{Z_{2}R}\right)^{2}}{1 + \left(\dfrac{A_{1}m}{Z_{1}R}\right)^{2}}} \
   \frac{J_{L1}(R_{L1})}{J_{L2}(R_{L2})} = \notag \\
   & = \mathcal{M}(R;A_{1}/Z_{1},A_{2}/Z_{2},\phi) \ \frac{J_{L1}(R_{L1})}{J_{L2}(R_{L2})},
\end{align}
where the subscripts 1 and 2 identifies $p_{1}$ and $p_{2}$, respectively.
The ratio of the modulated fluxes is the product of two terms: $\mathcal{M}$, which depends on the $A/Z$ difference between the two species; and the ratio of the interstellar spectra.
This last term is implicitly dependent on the $A/Z$ difference because each LIS is computed at the corresponding $R_{L}$, which is a function of $A/Z$.

\begin{figure}[h!]
   \centering
   \includegraphics[width=0.49\textwidth]{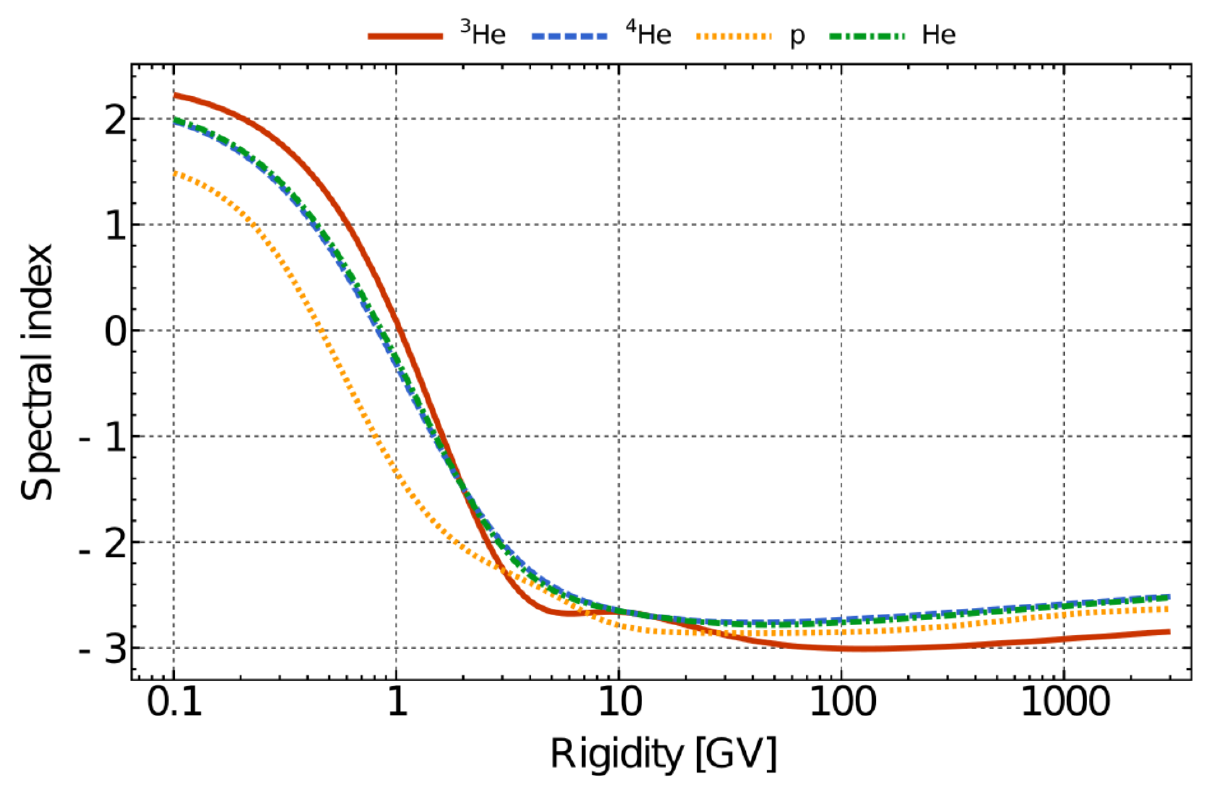}
   \includegraphics[width=0.49\textwidth]{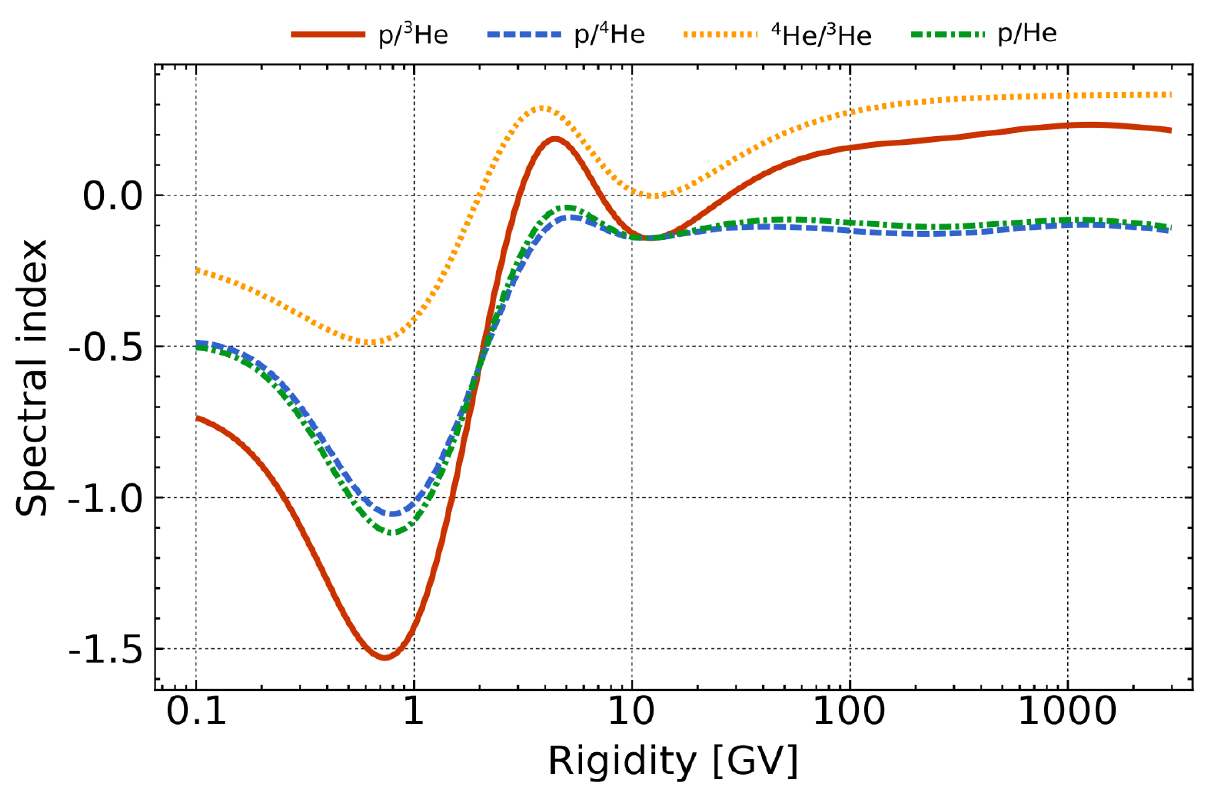}
   \caption{
      \textbf{Left.} Spectral index of the p, \He{3}, \He{4}, and He = \He{3} + \He{4} LIS, as function of rigidity.
      \textbf{Right.} Spectral index of the p/\He{3}, p/\He{4}, \He{4}/\He{3} and p/He LIS ratio, as function of rigidity.
   }
   \label{fig:LIS-spectral-index}
\end{figure}

In order to disentangle the mass-to-charge ratio dependence from the LIS shape dependence, we separately study the cases: (a) $A_{1}/Z_{1} = A_{2}/Z_{2}$ and $J_{L1}(R) \neq J_{L2}(R)$; (b) $A_{1}/Z_{1} \neq A_{2}/Z_{2}$ and $J_{L1}(R) = J_{L2}(R)$.
In the next two sections, we discuss these two cases with the p/He ratio as example, using the p, \He{3}, and \He{4} LIS defined in \cite{bib:corti19:model}.
For reference, Figure \ref{fig:LIS-spectral-index} shows the spectral index of the various LIS, defined as $\Gamma_{L}(R) = \dd\log J_{L}(R) / \dd\log R$, and the spectral index of the LIS ratio.
\section{LIS dependence}
If $A_{1}/Z_{1} = A_{2}/Z_{2} = A/Z$ and $J_{L1}(R) \neq J_{L2}(R)$, then $\mathcal{M} = 1$ and $R_{L1} = R_{L2} = R_{L}$.
Thus we have $J_{1}(R)/J_{2}(R) = J_{L1}(R_{L})/J_{L2}(R_{L})$, \ie\ the modulated ratio measured at $R$ depends only on the interstellar spectra ratio computed at the same rigidity $R_{L}$.
The time dependence of the modulated ratio is due to the $\phi(t)$ dependence of $R_{L}$.
The time derivative of the modulated ratio is:
\begin{equation}
   \frac{\dd}{\dd t}\left[ \frac{J_{1}(R)}{J_{2}(R)} \right] = \frac{J_{1}(R)}{J_{2}(R)} \ \Gamma_{\!\!r}(R_{L}) \ \frac{\phi + R\sqrt{1+(Am/ZR)^{2}}}{R_{L}^{2}} \ \frac{\dd\phi}{\dd t},
\end{equation}
where $\Gamma_{\!\!r} = \Gamma_{L1} - \Gamma_{L2}$ is the spectral index of the LIS ratio $J_{L1}/J_{L2}$.
We immediately see that the modulated ratio is correlated or anti-correlated with the change in solar activity ($\dd\phi/\dd t$) depending on whether $\Gamma_{\!\!r} > 0$ or $\Gamma_{\!\!r} < 0$, as the other factors multiplying the time derivative of the modulation potential are always positive.

\begin{figure}[h!]
   \centering
   \includegraphics[width=\textwidth]{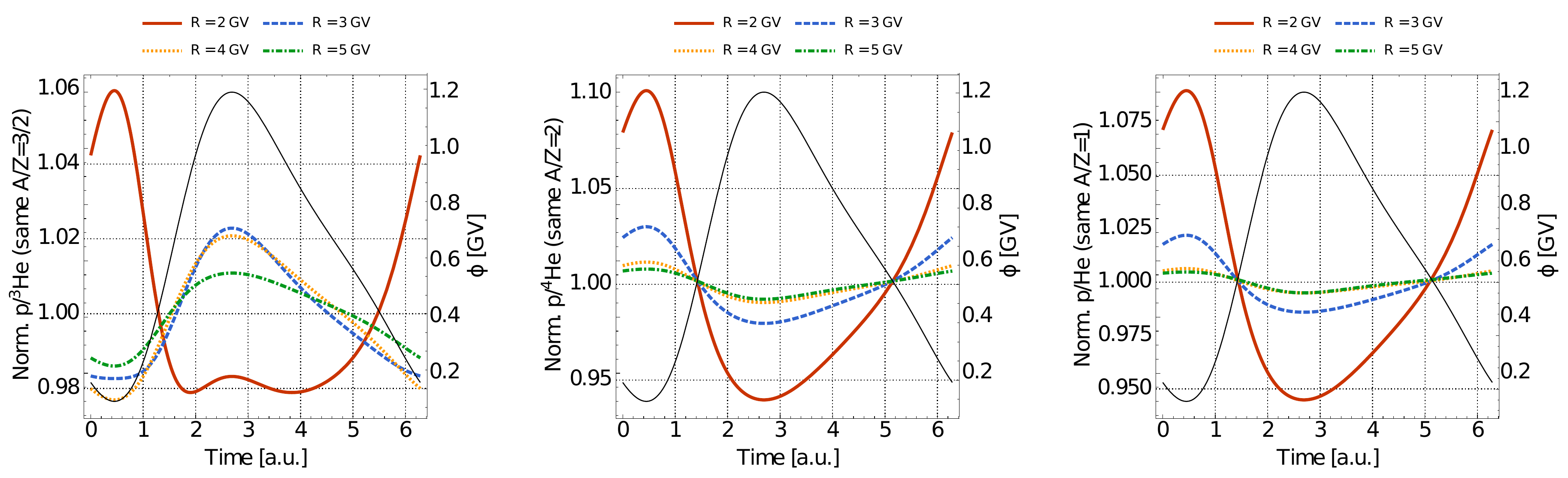}
   \caption{
      Time dependence of the modulated ratio (left axis), normalized with respect to the average value, between 2 GV and 5 GV (colored thick lines), assuming for both species the same $A/Z$, specified in the axis label, and the proper LIS (p LIS for p, \He{3} LIS for \He{3} LIS, etc).
      For reference, the time dependence of the modulation potential, $\phi(t)$, is shown as a thin black line (right axis).
   }
   \label{fig:LIS-dependence}
\end{figure}

For the sake of visualization, here and in the following sections, we simulate a time-dependent modulation potential, $\phi(t) = 0.1\ \mathrm{GV} + 0.55\ \mathrm{GV}\ [1 - \cos(t - 0.5\cos(t)) ]$, reproducing a solar cycle with a fast rising phase and slow decreasing phase.
Figure \ref{fig:LIS-dependence} shows the time dependence of p/\He{3} (left), p/\He{4} (center), and p/He (right), between 2 GV and 5 GV, assuming the same $A/Z$ for all species, but the proper LIS for each species,
\eg\ the p/\He{3} ratio has been computed with $A/Z=3/2$, while using the p LIS for p and the \He{3} LIS for \He{3}.
For reference, $\phi(t)$ is also shown as a black line.

Regarding p/\He{3}, we notice that at 2 GV the modulated ratio is anti-correlated with $\phi(t)$, except during the period of the solar maximum, where the ratio becomes correlated with $\phi(t)$.
This is due to the fact that $\Gamma_{\!\!r}\,$ is evaluated at $R_{L}$.
From Figure \ref{fig:LIS-spectral-index}, we see that $\Gamma_{\!\!r}$ is negative below 3 GV and positive between 3 GV and 7 GV.
If we compute $R_{L}$ for $R = 2$ GV and a varying $\phi$, we find that for $\phi > 0.67$ GV, $R_{L} > 3$ GV, so that $\Gamma_{\!\!r}\,$ switches sign during the solar maximum, explaining the change in time behavior of the modulated p/\He{3} at 2 GV at the peak of solar activity.
At the other rigidities shown in the figure, $\Gamma_{\!\!r}(R_{L})$ is always positive, so p/\He{3} is always correlated with the phase of the solar cycle.
The same reasoning applies for p/\He{4} and p/He, for which $\Gamma_{\!\!r}\,$ is always negative.
In particular, the time dependence of p/He is very similar to the one of p/\He{4} because \He{3} is at most 20\% of He, so it does not contribute too much to the time variation of p/He.

\section{Mass-to-charge ratio dependence}
If $A_{1}/Z_{1} \neq A_{2}/Z_{2}$ and $J_{L1}(R) = J_{L2}(R) = J_{L}(R)$, then we find:
\begin{equation}
   \frac{J_{1}(R)}{J_{2}(R)} = \mathcal{M}\ \frac{J_{L}(R_{L1})}{J_{L}(R_{L2})} \approx \mathcal{M} \left[ 1 - \frac{\delta R_{L}}{R_{L2}}\ \Gamma_{\!L}(R_{L2}) \right] + O\left( \delta^{2} R_{L} \right),
\end{equation}
where $\delta R_{L}= R_{L2}-R_{L1}$, and we used a Taylor expansion of $J_{L}(R_{L1})/J_{L}(R_{L2})$ around $R_{L1} = R_{L2} - \delta R_{L}$.
The expansion is justified by the fact that $\delta R_{L}$ above 2 GV is less than 0.1 GV and 0.2 GV for p/\He{3} and p/\He{4}, respectively, so that the approximate value is within 0.3\% and 1\% from the true value for p/\He{3} and p/\He{4}, respectively.
Since $\delta R_{L}/R_{L}$ is less than 5\% above 2 GV, the dependence on the LIS is suppressed, and the time behavior is basically all due to the factor $\mathcal{M}$.
The time derivative of $\mathcal{M}$ is:
\begin{equation} \label{eqn:M-time-deriv}
   \frac{\dd\mathcal{M}}{\dd t} = \left[
   3R^{2} \left( \frac{1/\beta_{2} + \phi/R}{R_{L2}^{2}} - \frac{1/\beta_{1} + \phi/R}{R_{L1}^{2}} \right) + 
   \frac{1/\beta_{2} - 1/\beta_{1}}{(1/\beta_{2} + \phi/R)(1/\beta_{1} + \phi/R)}
\right] \frac{\mathcal{M}}{R}\frac{\dd\phi}{\dd t},
\end{equation}
where $1/\beta_{i} = \sqrt{1+(A_{i}m/Z_{i}R)^{2}}$.
The modulated ratio is thus correlated or anti-correlated with the change in solar activity depending on the sign of the factor in square brackets.
It is easily shown that this factor is always positive for p/\He{3} and p/\He{4} above 2 GV.
%

\begin{figure}[h!]
   \centering
   \includegraphics[width=\textwidth]{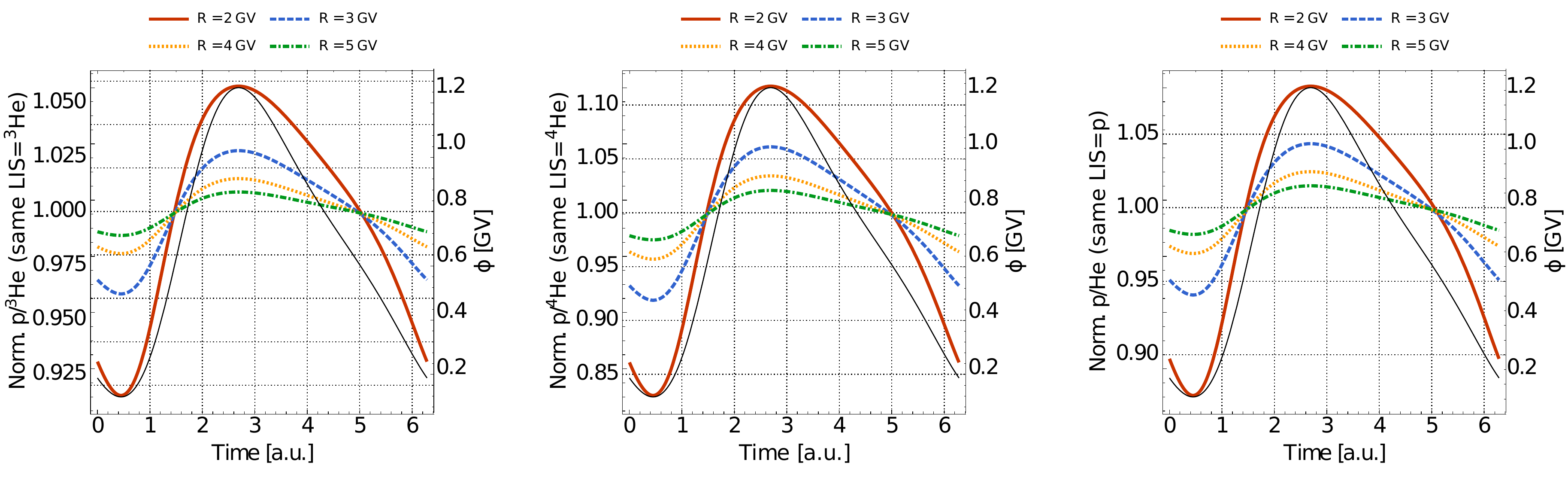}
   \caption{
      Time dependence of the modulated ratio (left axis), normalized with respect to the average value, between 2 GV and 5 GV (colored thick lines), assuming for both species the same LIS, specified in the axis label, and the proper $A/Z$ (1 for p, 3/2 for \He{3}, etc).
      For reference, the time dependence of the modulation potential, $\phi(t)$, is shown as a thin black line (right axis).
   }
   \label{fig:AZ-dependence}
\end{figure}

Figure \ref{fig:AZ-dependence} shows the time dependence of p/\He{3} (left), p/\He{4} (center), and p/He (right), between 2 GV and 5 GV, assuming the same LIS for all species, but the proper $A/Z$ for each species,
\eg\ the p/\He{3} ratio has been computed with the \He{3} LIS, while using $A/Z=1$ for p and $A/Z=3/2$ for \He{3}.
As expected, all the modulated ratios are correlated with $\phi(t)$.
Let us note also that the time behavior does not change too much when using different LIS, as a consequence of $\Gamma_{L}$ being multiplied by a small number.

\section{Conclusions}
\begin{figure}[h!]
   \centering
   \includegraphics[width=\textwidth]{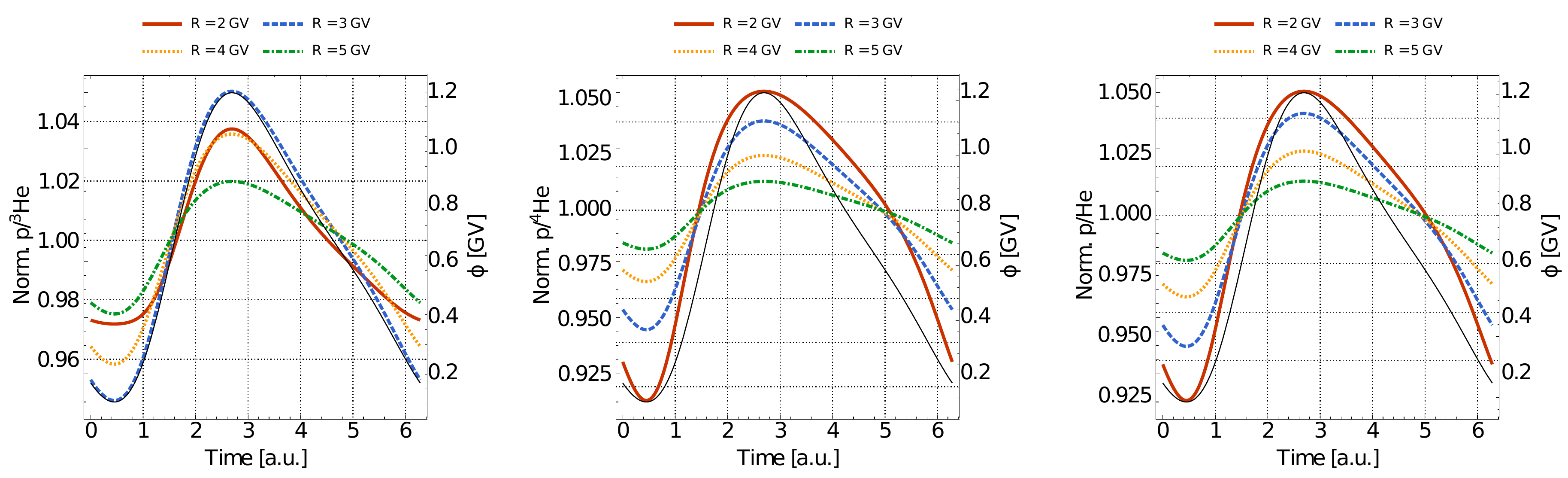}
   \caption{
      Same as Figure \ref{fig:p-He-norm}, but using the proper LIS and $A/Z$ for each species.
   }
   \label{fig:p-He-norm}
\end{figure}

In this work, we examined the dependence of the time variation of the flux ratio of two GCR species on their mass-to-charge ratio, $A/Z$, and LIS, according to the force-field approximation, in order to better understand the origin of the long-term decrease in time of the p/He ratio measured by AMS during the descending phase of solar cycle 24.
We focused our study on rigidities above 2 GV, which are pertinent to the energies observed by AMS.
For species with the same $A/Z$, but different LIS, the modulated ratio is anti-correlated (correlated) with the phase of the solar cycle if the spectral index of the LIS ratio is negative (positive).
In particular, we find that the differences in LIS between p, \He{3} and \He{4} induce a behavior of the p/He ratio at 2 GV opposite to what AMS observes, \ie\ an increase in time after solar maximum, since the spectral index of the p/He LIS ratio is always negative.
This suggests that the different LIS did not play a major role in the time variation seen by AMS.
For species with the same LIS, but different $A/Z$, the modulated ratio is instead always correlated with the phase of the solar cycle above 2 GV, reproducing a behavior similar to the observations.
Indeed, if we compute the p/He ratio using for each species the proper LIS and $A/Z$ (Figure \ref{fig:p-He-norm}), we obtain basically the same result as in Figure \ref{fig:AZ-dependence}, meaning that the different mass-to-charge ratio is the most probable cause of the time variation measured by AMS.

\acknowledgments
This work has been supported by: National Science Foundation Early Career under grant (NSF AGS-1455202); Wyle Laboratories, Inc. under grant (NAS 9-02078); NASA under grant (17-SDMSS17-0012).

\end{document}